\documentclass[prl,twocolumn,showpacs,preprintnumbers,amsmath,amssymb]{revtex4}

\usepackage{graphicx}
\usepackage{dcolumn}
\usepackage{bm}

\begin{document}


\title{Magic-wave-induced $^1\!S_0$$-$$^3\!P_0$ transition
in even isotopes of alkaline-earth-like atoms }

\author{Vitaly D. Ovsiannikov }
  \email{ovd@phys.vsu.ru}
 \affiliation{Physics Department, Voronezh State University,
 Universitetskaya pl. 1, 394006, Voronezh, Russia}
\author{Vitaly G. Pal'chikov }%
 \email{vitpal@mail.ru}
\affiliation{Institute of Metrology for Time and Space at National
Research Institute for Physical--Technical and Radiotechnical
Measurements, Mendeleevo, Moscow Region, 141579
Russia}%
\author{Alexey V. Taichenachev  and Valeriy I. Yudin }
\email{llf@laser.nsc.ru} \affiliation{Institute of Laser Physics SB
RAS,~Lavrent'ev~Avenue~13/3,~Novosibirsk~630090,~Russia
\\ Novosibirsk State University, Pirogova st. 2,~Novosibirsk~630090, Russia
}%
\author{Hidetoshi Katori and Masao Takamoto}
 \email{katori@amo.t.u-tokyo.ac.jp}
\affiliation{Department of Applied Physics, School of Engineering,
The University of Tokyo, Bunkyo-ku, Tokyo 113-8656, Japan
}%

\date{\today}

\begin{abstract}
The circular polarized laser beam of the ``magic'' wavelength may
be used for mixing the $^3\!P_1$ state into the long-living
metastable state $^3\!P_0$, thus enabling the strictly forbidden
$^1\!S_0$$-$$^3\!P_0$ ''clock'' transition in even isotopes of
alkaline-earth-like atoms, without change of the transition
frequency. In odd isotopes the laser beam may adjust to an optimum
value the line width of the "clock" transition, originally enabled
by the hyperfine mixing. We present a detailed analysis of various
factors influencing resolution and uncertainty for an optical
frequency standard based on atoms exposed simultaneously to the
lattice standing wave and an additional "state-mixing" wave,
including estimations of the ``magic" wavelengths, Rabi
frequencies for the ''clock'' and state-mixing transitions, ac
Stark shifts for the ground and metastable states of divalent
atoms.
\end{abstract}

\maketitle

Extremely narrow atomic line corresponding to a strictly forbidden
$^1\!S_0$$-$$^3\!P_0$ transition between ground and metastable
states of alkaline-earth-like atoms (such as Mg, Ca, Sr, Yb, Zn,
Cd), currently considered as worthwhile candidates for an optical
frequency standard, may be observed either on free odd isotopes
\cite{TK03, THHK05, HBOF05} or on even isotopes in external fields
\cite{HCW05,SAI05,TYO06,BHO06}. The mixing of the $^3\!P_1$ and
$^3\!P_0$ states by the hyperfine interaction in the odd isotopes
and by an external field in the even isotopes is the basic effect
which removes the general selection-rule restrictions on the
$0$$-$$0$ radiation transition. Intensive investigations of even
alkaline-earth-like isotopes during the last few years were
stimulated by a possibility to design a new frequency standard
based on an oscillator with the record high quality $Q$-factor. In
all the methods based on interrogation of the strongly forbidden
$^1\!S_0$$-$$^3\!P_0$ transition in the even isotopes embedded
into an optical lattice, engineered so as to equalize the upper-
and lower-level Stark shifts, some additional radiation
\cite{HCW05,SAI05} or static \cite{TYO06,BHO06} fields were
applied.

In this article, we propose to use a circularly (elliptically)
polarized wave of the ``magic'' wavelength $\lambda_{mag}$
(corresponding to the so-called ``Stark-cancellation'' regime, see
e.g. \cite{TK03, KTP03}), in addition to the optical lattice
field, in order to mix the $^3\!P_1$ state to the $^3\!P_0$ state.
Since in even isotopes the nuclear momentum equals zero, both the
initial and the final states of the frequency standard transition
(the ``clock'' transition) have zero total momenta, without
hyperfine structure splitting and without antisymmetric and tensor
increments to the ac dipole polarizabilities and to the Stark
effect. This makes the Stark shift of the upper and lower levels
independent of polarization of external fields. Meanwhile the
circular polarization of a laser wave allows for the second-order
dipole-dipole mixing of the $^3\!P_1$ state to the metastable
$^3\!P_0$ state, which is strictly forbidden for the linear
polarization.

So, the role of the optical lattice field consists in trapping
neutral atoms effectively free from collisions and Doppler effect
(Lamb-Dicke regime) as well as from the light field perturbations
\cite{TK03}, whereas an additional beam of the magic frequency,
but with compulsory circular (elliptic) polarization, will enable
the strictly forbidden radiation transitions via mixing the
$^3\!P_1$ state to the metastable $^3\!P_0$ state. The two waves
may be generated by one and the same laser or be completely
independent, each properly adjusted to some particular conditions.
So they may have different intensities, polarizations, wave
vectors and even different wavelengths, subject, however, to the
Stark-cancellation regime. In contrast with other methods
\cite{HCW05}--\cite{BHO06}, in this approach the atoms are exposed
to only the magic-wavelength radiation, no additional ac or dc
field is used and therefore no additional shift of the clock
frequency can arise.

The origin of the laser radiation-induced mixing consists in the
possibility of the second-order dipole transition between the
$^3\!P_1$ state and the metastable $^3\!P_0$ state in ac field of
a ``magic'' frequency $\omega$=$\omega_m$=$2\pi c/\lambda_{mag}$
($c$ is the speed of light) with a circular (elliptical)
polarization. To this end, together with the standing wave of the
optical lattice, a circularly polarized wave of the magic
frequency should be used, which we further consider as the running
wave with the electric field vector
\begin{equation} \label{Field}
\bm F_r(\bm r, t)=F_r{\rm Re}\left\{\bm e \cdot \exp [i(\bm
k\cdot\bm r-\omega_m t)]\right\},
\end{equation}
where $F_r$ is a real scalar amplitude, $\bm e$ is a complex unit
polarization vector, $\bm k=\bm n\omega_m/c$ is the wave vector
with the unit vector $\bm n$ which should have a non-zero
component at right angle to the optical lattice beam in order that
interrogation wave could travel along the lattice in compliance
with the Doppler-cancellation conditions (for simplicity, we
assume a 1D lattice here). The contribution of the $^3\!P_1$-state
wave function into the metastable $^3\!P_0$-state wave function is
determined by the ratio of the field-induced $^3\!P_0$$-$$^3\!P_1$
transition amplitude (Rabi frequency) $W_{10}$ to the
fine-structure splitting
$\Delta_{10}$$=$$E_{^3\!P_1}$$-$$E_{^3\!P_0}$. In the
nonrelativistic dipole approximation, the lowest non-vanishing
(second) order in $F_r$ amplitude (the atomic units are used in
this paper, if not otherwise indicated)
 \begin{equation} \label{W}
W_{10}=-\frac{F_r^2}{4\sqrt{6}}\, \xi\, \alpha^a_{^3\!P}(\omega_m)
 \end{equation}
is directly proportional to the circular polarization degree
$\xi=i(\bm n\cdot[\bm e\times\bm e^*])$ and to the antisymmetric
part $\alpha^a_{^3\!P}(\omega_m)$ of the$^3\!P_J$ triplet state ac
polarizability, which e.g. for the state with maximal total
momentum $J=L+S=2$ is (see \cite{MOR86, OPK06, DOZ93}):
\begin{eqnarray} \label{alpa1}
\alpha_{^3\!PJM}(\omega)&=&\alpha^s_{^3\!P}
(\omega)+\frac{M}{2J}\xi\alpha^a_{^3\!P}(\omega) \nonumber \\
&&-\frac{3M^2-J(J+1)}{2J(2J-1)}\, \alpha^t_{^3\!P}(\omega),
\end{eqnarray}
here $M$=$(\bm n\cdot\bm J)$ is the magnetic quantum number; the
superscripts ($s$) and ($t$) indicate the scalar and tensor parts
of the ac polarizability $\alpha_{^3\!PJM}(\omega)$.

Actually, the amplitude (\ref{W}) may be compared to the amplitude
of the hyperfine interaction, which mixes the states in the odd
isotopes  \cite{KTP03,OPK06}, or to the magnetic-field-induced
amplitude when the atoms experience the action of a magnetic
field, which may also be used for the $^3\!P_0$$-$$^3\!P_1$ state
mixing \cite{TYO06, BHO06}. Numerical computations carried out in
the single-electron approximation  with the use of the model
potential method for analytical presentation of the radial wave
functions \cite{MOR86,OPK06}, gave the numerical values of the
antisymmetric polarizabilities presented in Table \ref{T1} for Mg,
Ca, Sr, Yb, Zn and Cd atoms at the ``magic'' wavelength
corresponding to equal second-order ac Stark shifts $\Delta E(\,
^3\!P_0)=\Delta E(\, ^1\!S_0)= E ^{(2)}_L$ of the metastable and
ground states (the lattice depth)
  \begin{equation} \label{E2}
 E^{(2)}_L=-\frac14\alpha^s(\omega_m)F_L^2,
  \end{equation}
where $\alpha^s(\omega_m)=\alpha_{^1\!S_0}(\omega_m)
=\alpha_{^3\!P_0}(\omega_m)$ is the ac polarizability of the
``clock'' levels; $F_L$ represents the amplitude near the antinode
of the lattice standing wave, oscillating with the ``magic''
frequency $\omega_m$.

The Rabi frequency for the running-wave-induced transition
(\ref{W}) is directly proportional to the product of the wave
intensity $I_r$=$cF_r^2/8\pi$ and to the antisymmetric
polarizability $\alpha^a_{^3\!P}$, and may be presented in {\em
MHz}, as follows
 \begin{equation} \label{W2}
W_{10}=-0.01915\,\xi\alpha^a_{^3\!P}(\omega_m)I_r,
 \end{equation}
where $I_r$ is taken in $MW/cm^2$ and $\alpha^a_{^3\!P}$ in atomic
units. The value of $W_{10}$ determines the magnitude of the
coefficient
 \begin{equation} \label{a1}
a_1=\frac{W_{10}}{\Delta_{10}}
 \end{equation}
for the running-wave-induced contribution of the $^3\!P_1$ state
to the wave function of an atom initially (when the field
(\ref{Field}) is off) in the metastable $^3\!P_0$ state
 \begin{eqnarray} \label{psiL}
|\psi\rangle &=& |^3\!P_0\rangle +a_1  |^3\!P_1\rangle
 \nonumber \\
&& = |^3\!P_0\rangle +a_1\left
(a|^3\!P^{(0)}_1\rangle+b|^1\!P^{(0)}_1\rangle \right ),
 \end{eqnarray}
where the superscript $(0)$ indicates a pure $LS$-state. The
singlet-triplet mixing coefficients $a$ and $b$ in (\ref{psiL})
may be calculated using the ratio of the lifetimes $\tau
(\,^1\!P_1)$, $\tau (\,^3\!P_1)$ of singlet and triplet levels and
the wavelengths $\lambda(\,^1\!P_1-\,^1\!S_0)$,
$\lambda(\,^3\!P_1-\,^1\!S_0)$ of photons emitted in their
radiation decay, as follows:
 \begin{equation} \label{wL11}
\frac{b^2}{a^2}=\frac{\tau (\,
^1\!P_1)\lambda^3(\,^3\!P_1-\,^1\!S_0)}
{\tau(\,^3\!P_1)\lambda^3(\,^1\!P_1-\,^1\!S_0)}, \quad a^2+b^2=1.
 \end{equation}
For $I_r$ in $MW/cm^2$, $\Delta_{10}$ in $cm^{-1}$ and
$\alpha^a_{^3\!P}$ in atomic units the rate of the laser
field-induced radiation transition $^3\!P_0$$\to$$^1\!S_0$ may be
written as
 \begin{equation} \label{wL}
w=|a_1|^2w_{ic}=0.4080\cdot 10^{-12}
\left(\frac{\xi\alpha^a_{^3\!P}(\omega_m)I_r}
{\Delta_{10}}\right)^2w_{ic},
 \end{equation}
where $w_{ic}$=$1/\tau(\,^3\!P_1)$ is the field-free
$^3\!P_1$$\to$$^1\!S_0$ intercombination transition rate, the data
for which is presented in Table \ref{T2} (see e.g. \cite{PKRD01,
SJ02}).

\begin{table}
\caption{\label{T1} Numerical values of the ``magic'' wavelength
$\lambda_{mag}$, $^3\!P_1$$-$$^3\!P_0$ splitting
$\Delta_{10}=E_{^3\!P_1}-E_{^3\!P_0}$, anti-symmetric
polarizability $\alpha^a_{^3\!P}$ and the lattice-field-induced
second-order Stark shift (lattice depth) $ E^{(2)}_L$ for the
ground-state and metastable alkaline-earth-like atoms in the
optical lattice of the magic wavelength $\lambda_{mag}$ and
intensity $I_L$=10$\, kW/cm^2$. The transition matrix element
$W_{10}$ is given for the mixing-wave intensity $I_r=1\,MW/cm^2$.}
\begin{ruledtabular}
\begin{tabular}{ccccccc}
 Atom & $\lambda_{mag}$ &$\Delta_{10}$
 & $\alpha^a_{^3\!P}(\omega_m)$ & $E^{(2)}_L $ &$W_{10}/\xi$\\
 &$nm$&$cm^{-1}$ & $a.u.$ & $kHz$& $MHz$ \\
 \hline\\
 Mg & 432& 20.06 & 538.5 & $-49.3$ & $-10.3$\\
 Ca & 680 & 52.16 & $-1054$ & $-102$  & 20.2 \\
 Sr & 813.42\footnote{the experimentally determined value \cite{TK03,THHK05}}
 & 186.83 & $-1044$ &$ -116$ & 20.0 \\
 Yb & 759.35\footnote{the experimentally determined value \cite{BHO06}}
 & 703.57 & $-1084$ &$ -78.7$ & 20.8 \\
 Zn & 382 & 190.08 & $329.4$& $-21.3$ & $-6.31$  \\
 Cd & 390 & 542.1 & $390.6$ & $-25.8$ & $-7.48$ \\
\end{tabular}
\end{ruledtabular}
\end{table}

As follows from the data of Table \ref{T1}, at the intensity
$I_r$=0.5$\, MW/cm^2$, the absolute value of the
magic-wave-induced amplitude (\ref{W2}) may amount to 10 {\em MHz}
for atoms of Ca, Sr and Yb, that is equivalent to the amplitude
induced by a magnetic field of 1 $mT$ \cite{TYO06}. With account
of the data for the spin-orbit splitting of the lowest
(metastable) triplet state $^3\!P_J$ (see e.g. \cite{NIST}) the
admixture of the $^3\!P_1$ state in the $^3\!P_0$-state wave
function at these conditions does not exceed $10^{-5}$. Similar
estimates indicate that the $^1\!P_1$ singlet state admixture in
(\ref{psiL}) at these conditions is yet 4 to 5 orders smaller.
However, the $^3\!P_1$-state admixture may be sufficient to enable
the radiation transition between the ground and metastable states
and to amplify the magnitude of the $^3\!P_0$ level width by 7 to
9 orders (in comparison with a two-photon E1-M1 or three-photon E1
spontaneous radiation decay width \cite{OPK06}), up to several
$mHz$, making the clock transition $^1\!S_0$$\to$$^3\!P_0$ well
detectable, on the one hand, and on the other hand, retaining the
$^3\!P_0$ level width in bosonic atoms essentially smaller than in
fermionic.

Together with the radiative decay rate (\ref{wL}), the important
characteristic of the magic-wave-induced $^1\!S_0$$-$$^3\!P_0$
dipole transition, probed by the clock-frequency radiation, is the
amplitude (Rabi frequency) of the clock transition which after
integration in angular variables may be written as:
 \begin{equation} \label{Omega}
\Omega=\langle\, \psi|\hat v_p|\, ^1\!S_0\rangle=\beta
I_r\sqrt{I_p}\, \left(i\left[\bm e\times\bm e^*\right]\cdot \bm
e_p\right),
 \end{equation}
where $\hat v_p=\sqrt{I_p}(\bm e_p\cdot\bm r)$ is the Hamiltonian
of the dipole interaction between atom and probe field of
intensity $I_p$ and the unit polarization vector $\bm e_p$ which,
evidently, should be parallel to the running wave vector $\bm
k$$\propto$$\,i[\bm e$$\times$$\bm e^*]$, thus the maximal value
of $\Omega$ will be for orthogonal propagation to the probe beam.
So, in the case of a 1D optical lattice the Doppler-free
interrogation is possible when the probe beam propagates along the
lattice and is polarized along the mixing beam wave vector.

The coefficient $\beta$ includes all radial integrals of the
matrix element (\ref{Omega}) which may be presented in the units
of $mHz/(\sqrt{mW/cm^2}\cdot MW/cm^2)$, as follows:
\begin{equation} \label{beta}
\beta=204.9 \, \frac{\alpha^a_{^3\!P}(\omega_m)\langle
^1\!P_1^{(0)}|r|\, ^1\!S_0\rangle}{\Delta_{10}}\,b,
\end{equation}
with the antisymmetric polarizability and the radial part of the
dipole transition matrix element in atomic units, the splitting
$\Delta_{10}$ is in $cm^{-1}$. According to the calculated
numerical values of $\beta$ (see Table \ref{T2}), the Rabi
frequency in Sr and Yb atoms (\ref{Omega}) may achieve 0.3 {\em
Hz} for the field (\ref{Field}) intensity $I_r$=0.5$\, MW/cm^2$
and the probe field of $I_p$=10$\, mW/cm^2$.

\begin{table}
\caption{\label{T2}  Numerical values of the ``clock'' wavelength
$\lambda_c$, coefficients $\kappa_p^{(1)}$ and $\kappa^{(2)}$ of
linear in intensity of the probe field and quadratic in intensity
of the circularly polarized lattice wave and/or mixing wave Stark
shifts (\ref{domc}), the rate $w_{ic}$ of spontaneous
intercombination transition $^3\!P_1$$\to$$^1\!S_0$ and the
coefficient $\beta$ for the Rabi frequency (\ref{Omega}). The
number in parentheses determines the power of ten.}
\begin{ruledtabular}
\begin{tabular}{cccccc}
 Atom & $\lambda_{c}$ & $\kappa^{(1)}(\omega_c) $ & $ \kappa^{(2)}(\omega_m)$ & $w_{ic}$
 & $|\beta|$ \\
 &$nm$ & $\frac{mHz}{mW/cm^2}$& $\frac{Hz}{(MW/cm^2)^2}$ & $s^{-1}$
 & $\frac{mHz}{MW/cm^2\sqrt{mW/cm^2}}$\\ \hline\\
 Mg & 458& 4.27& -176 & 2.78(2) & 32.7  \\
 Ca & 660 & $-$4.50& -255  & 2.94(3)& $137.5$ \\
 Sr & 698 & $-$44.2& -61.5 & 4.70(4)& $176.9$  \\
 Yb & 578 & 24.5& -16.8 & 1.15(6) & $180.6$ \\
 Zn & 309 & 0.816 & -6.96 & 4.0(4) & 15.2  \\
 Cd & 332 & 23.0  & -10.3 & 4.17(5) & 22.6  \\
\end{tabular}
\end{ruledtabular}
\end{table}

In the Stark-cancellation regime, when the second-order ac Stark
shifts (\ref{E2}) of the clock levels are made equal to one
another, the clock frequency may be distorted by the
probe-field-induced quadratic ac Stark shift (linear in intensity
$I_p\propto F_p^2$) and the fourth-order ac Stark shifts of the
clock levels (quadratic in intensities $I_L\propto F_L^2$ and
$I_r\propto F_r^2$, correspondingly), induced by the lattice field
and the mixing wave, also including the bilinear in the
intensities $I_L$ and $I_r$ fourth-order correction. This shift
may be written as
\begin{eqnarray}
\Delta \omega_c&=&\kappa^{(1)}(\omega_c)I_p+\kappa^{(2)}(\bm
e_L,\omega_m) I_L^2 +\kappa^{(2)}(\bm e,\omega_m) I_r^2 \nonumber
\\ \label{domc} &+&\kappa^{(2)}(\bm e_L,\bm e,\omega_m) I_LI_r,
\end{eqnarray}
where the constant $\kappa^{(1)}(\omega_c)$ is determined by the
difference of the upper- and lower-level polarizabilities at the
clock-transition frequency $\omega_c=2\pi c/\lambda_c$. For
$\kappa^{(1)}$ in the units of $mHz/(mW/cm^2)$ the relation is
\begin{equation} \label{k2p}
\kappa^{(1)}(\omega_c)=-0.0469\left[\alpha_{^3\!P_0}(\omega_c)
    -\alpha_{^1\!S_0}(\omega_c)\right],
\end{equation}
where polarizabilities $\alpha_{^3\!P_0} (\omega_c)$ and
$\alpha_{^1\!S_0}(\omega_c)$ are in atomic units.

The coefficients  $\kappa^{(2)}$ are determined by the difference
of the clock-state hyperpolarizabilities at the ''magic''
frequency $\omega_m$ (similar to polarizabilities, the
hyperpolarizabilities for states with the total momentum $J$=0
include only scalar parts, which, however, depend on the wave
polarization vector $\bm e$ \cite{MOR86, DOZ93}),
\begin{eqnarray}
\kappa^{(2)}(\bm e,\omega_m)&=&-8.359\cdot 10^{-8} \nonumber \\
 \label{k4L} &\times& \left[\gamma_{^3\!P_0}(\bm e,\omega_m)
 -\gamma_{^1\!S_0}(\bm e,\omega_m)\right],
\end{eqnarray}
where $\kappa^{(2)}$ is in the units of $Hz/(MW/cm^2)^2$, the
hyperpolarizabilities $ \gamma_{^1\!S_0} (\bm e,\omega_m)$  and $
\gamma_{^3\!P_0} (\bm e, \omega_m)$ are in atomic units. Although
we assume the same ''magic'' frequency $\omega_m$ for the lattice
and running waves, the hyperpolarizabilities for the linear
polarization may differ essentially from those for the circular
polarization \cite{MOR86,DOZ93}, i.e. $\kappa^{(2)}(\bm
e_L,\omega_m) \neq\kappa^{(2)}(\bm e,\omega_m)$ for different
polarization vectors $\bm e_L$ and $\bm e$. The clock-level
hyperpolarizabilities determining the coefficient
$\kappa^{(2)}(\bm e_L,\bm e,\omega_m)$ of the interference term,
bilinear in the lattice-wave and running-wave intensities, depend
on the relative orientation (and the type -- linear or circular)
of polarization vectors $\bm e_L$ and $\bm e$. That is why the
fourth-order corrections from the both waves should be taken into
account together with the mixed bilinear correction
$\kappa^{(2)}(\bm e_L,\bm e,\omega_m)I_LI_r$.

The numerical estimates of $\kappa^{(1)}(\omega_c)$ and
$\kappa^{(2)}(\bm e, \omega_m)$ for the circular polarization of
the laser beam $\bm e$ are presented in table \ref{T2}. The
hyperpolarizabilities for the metastable $^3\!P_0$ levels of the
${\rm Mg}$, ${\rm Zn}$ and ${\rm Cd}$ atoms are complex values
with imaginary parts (determining the two-photon ionization width)
negligible in comparison with real parts. In estimating real parts
of the hyperpolarizabilities in Mg, Zn and Cd, we took into
account only the ``resonant'' terms, which may be determined by
the antisymmetric and tensor polarizabilities of the levels
\cite{DO84}.

The values of susceptibilities $\kappa^{(1)}$ and  $\kappa^{(2)}$
of Table \ref{T2} are the useful data to control the higher-order
corrections appearing when the probe-wave and running-wave
intensities increase. However, the strong dependencies  of
$\kappa^{(2)}$ both on the polarization and on the frequency
stimulate detailed investigations of the higher-order light shifts
in the close vicinity of the magic-wave frequency (see e.g.
\cite{BTB06, TYOP06}). Given the values $\kappa^{(1)}$ and
$\kappa^{(2)}$, all the combination of the probe-wave and
higher-order shifts (\ref{domc}) becomes controllable and may in
certain conditions be reduced to zero, using appropriate
intensities $I_L$, $I_r$, $I_p$, and polarization $\bm e_L$ and
$\bm e$ of the lattice and mixing waves.

In summary, we propose a new possibility to access the strongly
forbidden single-photon 0-0 transition of a bosonic
alkaline-earth-like atom supported by a magic-frequency wave with
circular (elliptic) polarization. This approach may be considered
as an alternative or an addition to the method of refs.
\cite{TYO06}, \cite{BHO06}, where a magnetic field is used. It
seems rather worthwhile for the lattice-based optical atomic clock
with the magic-wave-supported strongly forbidden transition of
even alkaline-earth-like isotopes. In the case of a 2D or 3D
lattice, the running wave (\ref{Field}) may be replaced by one of
the standing waves, which should be circular polarized  and
perpendicular to the probe laser beam. The circular polarization
of the standing wave coincides with that of the incident wave
whereas the amplitude near antinodes of the standing wave is twice
as big as the incident-wave amplitude. In this case the space
inhomogeneity of the mixing field should be taken into account,
but if atoms occupy the lowest vibration states of the lattice and
locate near antinodes then the field amplitude ''seen'' by an atom
is double what it is in the incident wave, therefore the
coefficients in the right-hand sides of equations (\ref{W}),
(\ref{W2}) and (\ref{Omega}) can be multiplied by 4. It means that
the Rabi frequencies (\ref{W2}) and (\ref{Omega}) for the given
laser input intensity  may become 4 times greater for the standing
wave in comparison with the running wave, in particular, in Sr and
Yb atoms, $W_{10}$=40$\, MHz$ and $\Omega$=1.2$\,Hz$ for
$I_r$=0.5$\, MW/cm^2$ and $I_p$=10$\, mW/cm^2$.

VDO acknowledges the support from the CRDF (USA) and MinES of
Russia (BRHE program, award VZ-010), AVT and VIYu were supported
by RFBR (grants 05-02-17086, 05-08-01389, 07-02-01230,
07-02-01028) INTAS-SB RAS (grant 06-1000013-9427), and by
Presidium SB RAS.


\begin{thebibliography}{99.}
\bibitem{TK03}
 M. Takamoto and H. Katori, Phys. Rev. Lett. \textbf{91}, 223001
 (2003).
\bibitem{THHK05}
 M. Takamoto, F.-L. Hong, R. Higashi and H. Katori, Nature
 \textbf{435}, 321 (2005).
\bibitem{HBOF05}
 C. W. Hoyt, Z. W. Barber, C. W. Oates, T. M. Fortier, S. A. Diddams,
 and L. Hollberg, Phys. Rev. Lett. \textbf{95}, 083003 (2005).
\bibitem{HCW05}
 T. Hong, C. Cramer, W. Nagourney, and E. N. Fortson, Phys. Rev. Lett.
 \textbf{94}, 050801 (2005).
\bibitem{SAI05}
 R. Santra, E. Arimondo, T. Ido, C. H. Greene, and J. Ye,
 Phys. Rev. Lett. \textbf{94}, 173002 (2005).
\bibitem{TYO06}
A. V. Taichenachev, V. I. Yudin,  C. W. Oates, C. W. Hoyt, Z. W.
Barber, and L. Hollberg, Phys. Rev. Lett. \textbf{96}, 083001
(2006).
\bibitem{BHO06}
Z. W. Barber, C. W. Hoyt, C. W. Oates,  L. Hollberg, A. V.
Taichenachev, and V. I. Yudin, Phys. Rev. Lett. \textbf{96},
083002 (2006).
 \bibitem{KTP03}
 H. Katori, M. Takamoto, V.G. Pal'chikov, and V.D. Ovsiannikov,
 Phys. Rev. Lett. \textbf{91}, 173005  (2003).
 \bibitem{MOR86}
N. L. Manakov, V. D. Ovsiannikov, and L. P. Rapoport, Phys. Rep. \textbf{141}, 319 (1986).
\bibitem{OPK06}
 V. D. Ovsiannikov, V. G. Pal'chikov, H. Katori, and M. Takamoto,
 Quantum Electronics \textbf{36}, 3 (2006).
 \bibitem{DOZ93}
 V.A. Davydkin, V.D. Ovsiannikov, and B.A. Zon, Laser Physics
 \textbf{3}, 449 (1993).
 \bibitem{PKRD01}
S. G. Porsev, M. G. Kozlov, Y. G. Rakhlina and  A. Derevianko,
Phys. Rev. A \textbf{64}, 012508 (2001).
 \bibitem{SJ02}
I. M. Savukov and W. R. Johnson, Phys. Rev. A \textbf{65}, 042503 (2002).
  \bibitem{NIST}
  NIST Atomic Spectra Database Levels Data,
  http://Physics.nist.gov
    \bibitem{DO84}
 V. A. Davydkin and V. D. Ovsiannikov,  J. Phys. B:
At. Mol. Opt. Phys. \textbf{17}, L207 (1984).
 \bibitem{BTB06}
 A. Brusch, R. Le Targat, X. Baillard, M. Fouche,
 and P. Lemonde, Phys. Rev. Lett. {\bf 96}, 103003 (2006).
 \bibitem{TYOP06}
A. V. Taichenachev, V. I. Yudin,  V. D. Ovsiannikov and V. G.
Pal'chikov, Phys. Rev. Lett. \textbf{97}, 173601 (2006).
\end{thebibliography}
\end{document}